\documentclass[twocolumn,showpacs,aps,prl]{revtex4}
\usepackage{graphicx}
\usepackage{amsfonts}
\usepackage{amsmath}
\usepackage{amssymb}
\usepackage{multirow}
\usepackage{braket}
\usepackage[usenames,dvipsnames]{color}

\begin{document}

\title{Tuning the mobility of a driven Bose-Einstein condensate via diabatic Floquet bands}
\author{Tobias~Salger$^1$, Sebastian~Kling$^1$, Sergey~Denisov$^2$, Alexey~V.~Ponomarev$^2$, Peter~H\"anggi$^2$, and Martin~Weitz$^1$}
\affiliation{$^{1}$Institut f\"ur Angewandte Physik der Universit\"at Bonn, Wegelerstrasse 8, 53115 Bonn, Germany\\
$^{2}$Institute of Physics, University of Augsburg, Universit\"{a}tstrasse 1, 86159 Augsburg, Germany}
\date{\today}
\begin{abstract}

We study the response of ultracold atoms to a weak force in the
presence of a temporally strongly modulated optical lattice
potential. It is experimentally demonstrated that the strong
ac-driving allows for a tailoring of the mobility of a
 dilute atomic Bose-Einstein condensate with the
atoms moving ballistically either along or against the direction
of the applied force. Our results are in agreement with a
theoretical analysis of the Floquet spectrum of a model system,
thus revealing the existence of diabatic Floquet bands in the
atom's band spectra and highlighting their role in the
non-equilibrium transport of the atoms.

\end{abstract}

\maketitle

Experiments with ultra-cold atomic gases in periodically modulated
lattice potentials have shown that these systems represent an
attractive testing ground to explore non-equilibrium quantum
states ~\cite{raizen, gemelke}. Along these lines, the suppression
of tunneling in ac-driven optical lattices was experimentally
observed,  both in the single-particle ~\cite{arimondo} and in the
many-body~\cite{bloch2} regimes. Other experiments have recently
reported the simulation of gauge fields for neutral and spinless
atoms~\cite{sengstock}.

Common knowledge in non-equilibrium
statistical physics \cite{non} indicates that not only the state
of a system shifted from equilibrium  but also its response to
external perturbations can differ substantially from those
exhibited by the system at rest. The equilibrium response of one
of the simplest quantum models, a
particle placed in a stationary periodic potential, when exposed
to a weak constant force $F$ is well understood: On a short-time
scale, $t \ll 2\pi \hbar/F L = T_B$, where $L$ is the period of
the potential, the particle reacts by moving with the velocity
determined by the local slope of the ground-band. On larger time
scales, $t \geq 2\pi \hbar/F L$, the response evolves into the
celebrated phenomenon of Bloch oscillations \cite{bl-zener,
gustav}. Weak periodic modulations of the potential can
change the response. It has been demonstrated that these
can propulse the particle over many Brillouin zones, thus rectifying
Bloch oscillations into ballistic transport  \cite{super_b}.
This observation, along with the results reported in the
above-cited works \cite{arimondo, bloch2, sengstock}, are well
explained by assuming that the temporal modulations do not push
the particle outside the ground band.  The single-band approach is
justified as long as the driving amplitude remains small, so the
observed effects can be attributed to a  modulation-induced renormalization of the
potential \cite{renorm, holt0}. Strong
driving, however, intermingles eigenstates of the stationary
system and sculptures a new spectrum of time-dependent `dressed
states` ~\cite{dressed, holt2}, so that the system dynamics no
longer fits the perturbative picture \cite{restr}. This idea has
been exploited in recent experiments with coherent quantum
ratchets~\cite{ketzmerick1,den,quantum_ratchet} and was also used to
create new topological states ~\cite{topology}.

A distinct feature of strongly driven quantum systems is the presence
in their Floquet spectra \cite{fei,floquet}
of so-called \textit{diabatic bands} \cite{takami, kolovsky}. The states  belonging to these bands remain near isolated
from the rest of system Hilbert space upon parameter variations. In the quasi-classical limit diabatic bands are also
called 'regular bands' because
of the correspondence between the band eigenstates  and regular invariant  manifolds of the  system in the
classical limit \cite{kolovsky}.
It is possible to detect a diabatic band of a semi-classical system by populating it with an initial wave packet located in the corresponding
region of the classical phase space. Evidently, this recipe no longer applies for  the systems operating in the deep quantum limit.

Here we show that in this limit diabatic Floquet states reveal
their presence via a strong ballistic-like response of a driven
quantum particle to a weak net force.
Upon changing  the modulation parameters, we populated different
diabatic bands and switched between regimes of positive and
negative responses, i.e. the particle then moves against the
applied bias. We measured the mobility \cite{mobility} of a dilute
atomic rubidium Bose-Einstein condensate (BEC) in an ac-driven
optical potential. Good agreement between the experimental results
and our theoretical model is obtained.

\textit{Model.} Consider  a  particle with mass $M$ moving
in a time- and space-periodic potential $U(\hat{x},t) = V(\hat{x})  A(t)$,
where the periodic functions
$V(x+L)=V(x)$ and  $A(t+T)=A(t)$ possess  the periods $L$ and
$T = 2\pi/\omega$, respectively. In addition, let the particle be exposed
to a tunable net bias $F$. The corresponding Hamiltonian then reads:
\begin{equation}\label{Eq:ham}
\hat{H} = \hat{p}^{2}/2M + V(\hat{x}) A(t) - F \hat{x}.
\end{equation}
The Hamiltonian of the non-driven system, i.e. when $A(t)\equiv 1$
and $F = 0$, is a spatially-periodic operator and its reciprocal
space is spanned by the Bloch bands, $E_{n}(\kappa)$, $\kappa \in
[-\pi/L, \pi/L]$,  $n = 1, 2, ...$ . By assuming that the initial
state is localized at the point with quasimomentum $\kappa = 0$,
the action of a bias $F$ can be considered as a linear ramp of the
quasimomentum, $\kappa(t) = Ft/\hbar$. On a time scale $t \ll
T_B$, the response of the system to a weak bias is determined by
the band-curvature
 $\hbar^{-2}\partial^{2} E_{n}(k)/\partial \kappa^{2}$, i.e. by the effective mass   ~\cite{kittel}.
The curvature of the ground-state band is  small and  positive
near the center of the Brillouin zone. One could, in principle,
obtain a variety of  mobility responses by placing the particle into different
excited bands, however the exponentially small
splitting between neighboring bands makes this idea less
practical. It is possible to prepare an initial atomic sample
outside of the center of the Brillouin zone, where the slope of
the ground-band is non-zero ~\cite{peik}; this would mean that the atoms were already
set into a slow motion in the absence of any bias.

\begin{figure}[t]
\center
\includegraphics[width=0.49\textwidth]{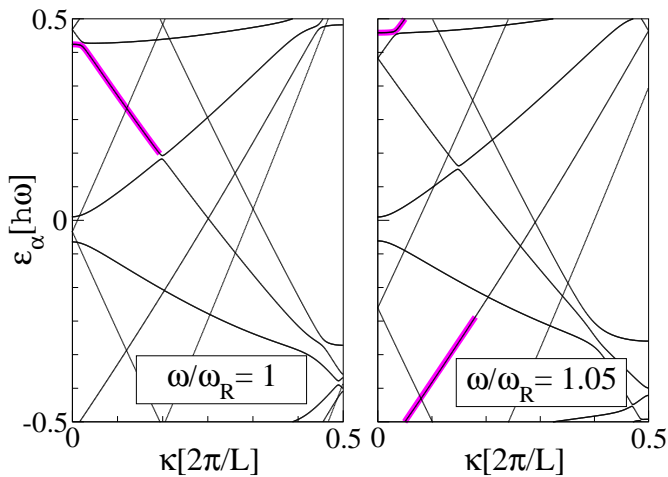}
\caption{(color online) Calculated Floquet band spectra of the
system, Eqs. (\ref{Eq:ham},\ref{eq:spatial}-\ref{eq:temporal}), for
typical experimental parameters. Only the first eight Floquet bands, maximally
overlapping with the Bloch band of the undriven potential, are
depicted. The heavy magenta lines mark the Floquet states with
largest overlap to the initial atomic state (see text), which here
transform themselves into ballistic diabatic bands after passing a
first, relatively broad, avoided crossing. A small change in the
driving frequency allows switching from a ballistic band with
negative velocity (left) to the symmetry-related band of positive
velocity (right). The time-reversal symmetry of the system
dynamics implies that the negative part of the  Brillouin zone is
a mirror image of the presented one. The used parameters are $V_1
= 0.352\times 16 E_r$, $V_2 = 0.11\times 16 E_r$, $A_1 = 0.66$,
$A_2 = 0.4$. The driving frequency is measured in units of
$\omega_R = 8\omega_r$ (see text). The parameters $\omega_r$ and
$E_r=\hbar \omega_r$ denote the recoil frequency and the recoil
energy, respectively.} \label{Fig:1}
\end{figure}

The dynamics of such a  time-modulated system
(\ref{Eq:ham}), $A(t) \neq const$, is more diverse. The
Hamiltonian is periodic in time and the solution of the
corresponding Schr\"{o}dinger equation for a given value of the
quasimomentum $\kappa$; i.e.,
\begin{equation}\label{Eq:floc}
[\hat{H}(\kappa, t) - i\hbar \partial_t] \Psi_{\kappa}(t)=0,~~~\hat{H}(\kappa, t+T) = \hat{H}(\kappa, t),
\end{equation}
can be obtained by solving the eigenvalue problem for the operator
that propagates the system over one period of
driving~\cite{floquet},
\begin{eqnarray} \nonumber
|\psi_{n,\kappa}(T)\rangle=\hat{U}(T)|\psi_{n,\kappa}(0)\rangle = ~~~~~~~~~~\\
~~~~~~~~~\exp(-i \epsilon_{n}(\kappa)t/\hbar)|\phi_{n,\kappa}(0)\rangle.
\label{Eq:floc_op}
\end{eqnarray}
The eigenfunctions obey the Floquet theorem,
$|\psi_{n,\kappa}(t)\rangle=\exp(-i \epsilon_{n}(\kappa)t/\hbar)$
$|\phi_{n,\kappa}(t)\rangle$, where Floquet states are
time-periodic, $|\phi_{n,\kappa}(t+T)\rangle=
|\phi_{n,\kappa}(t+T)\rangle$. The quasi-energies
$\epsilon_{n}(\kappa)$ are conventionally restricted to the
interval $[-\hbar \pi/T,\hbar \pi/T]$. Fig.~\ref{Fig:1} depicts
Floquet band spectra of the temporally periodically driven optical
lattice system, for typical experimental parameters and two
different drive frequencies. The spectra exhibit a complex,
web-like structure, which can be tailored by varying the
modulation $A(t)$. The finiteness of the quasi-energy range
creates a certain problem with the ordering of Floquet bands. The
concept of adiabatic following of a band looses its mathematical
rigor here because the set of avoided crossing points is dense
everywhere \cite{hone, holt2}. However, there exists no problem
with the concept of \textit{diabatic} following \cite{takami}. A
diabatic band can be obtained by moving through the parameter
space and connecting segments of different bands by ignoring all
avoided crossings met on the way  when the gap width of the
upcoming avoided crossing is below certain threshold
$\varepsilon$, see Refs. \cite{takami, kolovsky} for details. The
threshold is related to the velocity of the excursion through the
parameter space, which here is given by the strength of the bias,
$F$ \cite{vague}. The corresponding diabatic bands assume straight
lines,  running across the Brilloiun zone \cite{kolovsky,
ketzmerick}, cf. Fig.~1b.

Transport properties of the $n$-th Floquet state are characterized by the average velocity \cite{ketzmerick},
$\bar{\upsilon}_{n,\kappa}=\langle \upsilon_{n,\kappa}(t) \rangle_T$,
where $\upsilon_{n,\kappa}(t)$ is the instantaneous expectation value of the velocity operator, $ \hat{\upsilon} = (-i\hbar/M)\partial_x$,
and $\langle...\rangle_T$ denotes a time average over one period of the  driving \cite{frame}.
By virtue of the Hellmann-Feynman theorem \cite{fei,floquet},
the average velocity of the state is equal to the local slope of the corresponding Floquet band \cite{ketzmerick},
\begin{equation}\label{Eq:sambe}
\bar{\upsilon}_{n,\kappa} = \hbar^{-1}\partial\epsilon_{n} (\kappa) / \partial\kappa.
\end{equation}
Since diabatic bands have near constant velocities,
they are entitled  to be termed 'ballistic bands'.
The issue of symmetry is significant here: When the system Hamiltonian (\ref{Eq:ham}) is  invariant under  time-  or space-reversal, all
Floquet bands are flat at the center of the Brillouin zone, thus exhibiting   zero average velocities ~\cite{ketzmerick1,den}.
Formally, the presence of  a symmetry implies that the potential function $U(x,t) = A(t) V(x)$ remains invariant either under the
transformation $\hat{S}_t: t \rightarrow -t + \tau,~ x \rightarrow x + \chi$ or $\hat{S}_x: t \rightarrow t + \tau,~ x \rightarrow -x + \chi$,
where $\tau$ and $\chi$ are appropriate shift constants, see in ~\cite{den}.
In contrast to genuine Floquet bands, a diabatic band does need to
be periodic in quasi-momentum $\kappa$-space ~\cite{ketzmerick1},
and sometimes can wrap  the Brillouin zone several times before
meeting a broad  avoided crossing and loosing the diabaticity
property. This fact does not contradict the above symmetry
statement because the presence of  a symmetry only means that two
ballistic bands with opposite velocities cross each other at the
point $\kappa = 0$. Note however that such a 'crossing'
mimics  a narrow avoided crossing between the corresponding
pair of genuine Floquet bands.

Consider next the situation when at time $t=0$ a dilute and
de-localized cloud of ultracold atoms is loaded into the
temporally driven lattice potential. The initial atomic wave
function $|\varphi_0\rangle$ has the form of a wave-packet,
well-localized in quasimomentum at the point $\kappa = 0$. For our
calculations, we assume that the initial atomic wave-function is of
the form of the Bloch ground-state of the non-driven lattice
potential, which is experimentally reasonable.
We can order the Floquet states, $FS[n]$, according to their overlaps with
the initial state, i.e., $c_n = |\langle
\phi_{n,\kappa=0}(0)\lvert \varphi_0\rangle|^2$, $c_1 > c_2 >
...$. We find that, for our experimental parameters, the overlap
with  the first Floquet band is
$c_1\simeq 0.85$, and it is naturally to expect that the system mobility is
determined mainly by the properties of this band. We are interested in the situation where this state
transforms itself into a ballistic band outside the center of the
Brillouin zone, for example, due to a broad avoided crossing with
another Floquet state, as is the case for the parameters used in
Fig. 1. A  motion through the $\kappa$ space, induced by a weak
bias force, is slow enough for the system to remain on the band when passing through this AC.
We expect that the non-vanishing slope of the band beyond this point will
determine the mean particle velocity. In
general, the described scenario assumes that (i) a significant
population of the relevant band can be achieved and (ii) the
band group velocity is high. There is no general
recipe how to manage such a situation. There are, however, several
prerequisites serving as a guide. First, the system should operate
in  the deep quantum regime, ideally with only  few Floquet states
within the potential range. Otherwise, the initial wave function
will be distributed among several Floquet states of different
group velocities. Their joint contributions then yields a
inconclusive asymptotic response, while interference effects will
blur the finite-time response even more. Second, the modulation frequency
should be chosen close to the frequency of oscillations at the
bottom of the potential well, $\omega_R = 4\pi^{2}\hbar/L^{2}M$.
In numerical studies we have observed that the resonant modulation
typically produces at the vicinity of the initially populated
$FS[1]$-state at $\kappa=0$ two well-separated Floquet
states that are ballistic, with opposite velocities just outside
the center of the zone. By slightly adjusting the frequency of the
driving, one can then map the initially populated Floquet state
(which has a vanishing group velocity) into a ballistic state by
means of a broad avoided crossing between the $FS[1]$-state
and one of the states of the ballistic pair.

\begin{figure}[t]
\begin{tabular}{cc}
\includegraphics[width=0.21\textwidth]{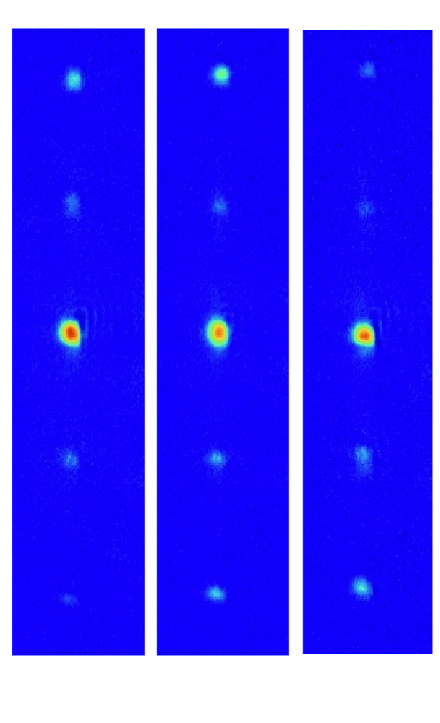}
\includegraphics[width=0.245\textwidth]{fig2b.eps}
\end{tabular}
\caption{(color online) (a) Time-of-flight (TOF) images of the
atomic cloud after $t_{exp}=26T$ periods of modulation for
$F=-0.0181~E_r/\lambda$ (left), $0$ (center), and
$0.0181~E_r/\lambda$ (right). The visible atomic diffraction
orders are $s=-2, ..., 2$ (from bottom to top). (b) Measured mean
velocity of the atomic cloud in the laboratory frame {\it vs.}
applied bias $F$. Note that the overall anti-symmetric response
behavior; i.e.,  $\bar v(F)= -\bar v (-F)$. The mean velocity of
the atomic cloud was calculated as $\bar{v}=\bar{p}/M_{Rb}$ with
$\bar{p}=2\hbar k\sum_s s |c_s|^2$, where $|c_s|^2$ denotes the
fraction of atoms in the $s$-th order momentum state, $|2s\hbar
k\rangle$, with $s = \pm 1, \pm 2$. The error bars show the
standard deviation of the mean value. The enlarged (red) data
points correspond to the TOF images. The solid lines depict the
results of numerical simulations of the theory. The initial state
was chosen in the form of a narrow Gaussian packet in the
quasimomentum space, with the center at $\kappa = 0$ and
dispersion $\sigma_\kappa = 0.04\hbar k$ ~\cite{sup}. The other
parameters are the same as in Fig.1.} \label{Fig:2}
\end{figure}

\textit{Experiment.} In our experiments we used an optical
potential, $U(x,t) = V(x) A(t)$, where $V(x)$ and $A(t)$ are of
the form~\cite{kolovsky, den, quantum_ratchet}
\begin{eqnarray}
\label{eq:spatial}
V(x) &=& V_1/2\cos(2kx) + V_2/2\sin(4kx)\,,  \\
\label{eq:temporal}
A(t)&=& A_1 \sin^{2}(\omega t/2) + A_2 \cos^{2}(\omega t)\,.
\end{eqnarray}
Here, $\lambda$ is the wavelength of the driving laser field and
$k=2\pi/\lambda$. Using the spatial lattice period, $L=\lambda/2$,
we arrive at a resonance frequency of $\omega_R = 8\omega_{r}$,
where $\omega_{r}=\hbar k^2/2M_{Rb}$ is the recoil frequency, and
$E_r=\hbar\omega_r$ denotes the recoil energy. Note that this
setup, although possessing a ratchet-like spatial profile
~\cite{tooth}, is perfectly {\it time-symmetric} since $A(-t) =
A(t)$ \cite{singleH}. Therefore, in the absence of an external bias force, $F=0$,
no  transport occurs \cite{den, rmp2009}, i.e.,  the average
current vanishes for any initial atomic wave-packet with zero
average velocity. We performed the experiments by loading a
Bose-Einstein condensate of rubidium atoms into a periodically
time-modulated optical  potential, of the form given by Eq.
(\ref{Eq:sambe}), to an external dc bias force.

We start the experimental sequence by preparing a
Bose-Einstein condensate of rubidium ($^{87}$Rb) atoms in a far
detuned optical dipole trap. During the next step the atoms are
allowed to freely expand ballistically for $2.5$ $ms$, which
reduces the atomic density. This leaves the atomic cloud with
essentially kinetic energy only, while the interaction energy is
strongly diminished. After then loading the atoms into an optical
lattice potential, a dc bias is realized by moving the lattice
potential with a constant acceleration. This acceleration
emulates a constant force in the co-moving frame~\cite{sup}. The
measured mean velocity of the atomic cloud versus the bias
strength $F$ is depicted  in Fig.~ 2b. For small values of $F>0$,
a negative response is clearly detectable, i.e. the average atomic
velocity is opposite to the applied bias force. We attribute this
to the population of the Floquet state marked by the heavy line in
the left panel of Fig. 1, which for positive quasimomentum values
has a negative group velocity. For larger absolute values of the
dc force, above roughly $|F|=0.02~E_r/\lambda$, the mobility again
becomes positive. This is attributed to the population of other
Floquet states when moving beyond the quasimomentum region
indicated by the heavy magenta line in Fig. 1. We have also
investigated the dependence of the atomic transport on the drive
frequency, see Fig. 3a. By changing the frequency to slightly
higher values, the atomic mobility can be tuned to a normal, i.e.,
positive response. This response is attributed to the overlap of
the $FS[1]$-state with the symmetry-related counterpart of the
previously involved ballistic band, as shown in the right panel of
Fig. 1.

\begin{figure}[t]
\includegraphics[width=0.45\textwidth]{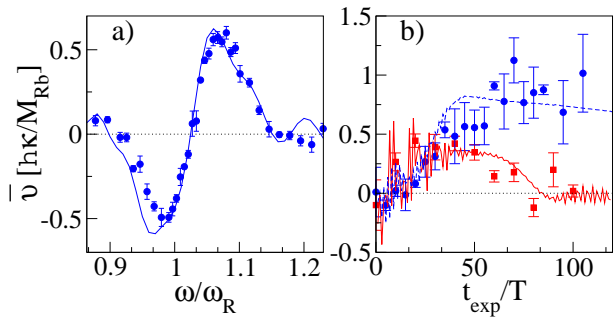}
\caption{(color online) (a) Average velocity of the atomic cloud in the laboratory frame as a function of
the modulation frequency $\omega$ (blue dots)
for a  bias strength $F=-0.0181~E_r/\lambda$ and exposition time $26T$; (b) Time evolution of the average cloud velocity for driving frequencies
$\omega/\omega_R=1.05$ (red squares) and $\omega/\omega_R=1.076$ (blue dots).
The solid lines depict the results of numerical simulations~\cite{sup}.
The error bars show the standard deviation of the mean.
The remaining parameters are the same as in Fig.~1.}
\label{Fig:3}
\end{figure}

An important issue is the stability of the mobility
response. Let us first discuss this issue in two different limits,
classical and quantum ones. In the classical
dissipationless limit, a stationary motion against a constant
bias is possible~\cite{bias} due to the existence of invariant
manifolds, i.e. transporting regular islands, periodic orbits, and
cantori ~\cite{Ham_chaos}, in the phase space of an ac-driven
Hamiltonian system~\cite{ketzmerick}. 
Diabatic bands can be viewed as the quantum counterpart  of
classical ballistic manifolds. However, this analogy is not exact.
In contrast to classical manifolds, diabatic  bands are not
completely isolated since any finite bias  sets these bands into a
contact with other states ~\cite{LZF}. In addition, the band can loose its diabatic property by encountering
a broad  avoided crossing, see left panel of Fig. 1.  However, the band  could run over a substantial
region of the $\kappa$-space -- or even perform several
revolutions around the Brillouin zone -- before this happens,
thus transforming the ballistic response into a long-lasting metastable phenomenon. In
order to experimentally investigate this issue, we have
measured the velocity of the atomic cloud as a function of the
exposition time for two different values of the driving frequency,
both of which produced  negative
responses. Fig.~3b presents the results of the
measurements.  For a modulation frequency $\omega/\omega_R=1.05$
(red dots), the mean momentum of the atomic cloud  initially increases,
reaching a broad maximum value of $0.4\hbar k$ after
around 30 modulation periods, and then starts to decrease again,
approaching zero at $t \approx 80T$. When slightly changing the
modulation frequency to $\omega/\omega_R=1.076$ (blue squares),
the observed atomic momentum increases to roughly $0.9\hbar k$.
This value is  maintained to much longer modulation times. This finding is in agreement with the
fact that the  structure of Floquet bands  is very sensitive to variations of
the system parameters and, correspondingly, the system response can be controlled
over a wide range.


\textit{Conclusions.} The experimental detection of diabatic Floquet states in a time-dependent driven Bose-Einstein condensate
opens several directions that are worth exploring further.
For example, there is the issue relating to the constructive role which decoherence may play for the stabilization
of the  asymptotic response, similar to what
has been observed with stochastic models operating in the classical regime~\cite{machura}. Another promising research avenue
is the implementation of the tunable dispersion relation, stemming  from the  Floquet
spectrum  of temporally-modulated  optical lattices, for  simulations of relativistic
physics effects with ultra-cold atoms ~\cite{Salger_rel,Esslinger}. Finally, the inclusion of interaction between
atoms of a dense condensate~\cite{bloch1,bbbb} may open a way to the detection of many-body diabatic Floquet states.

This work was supported by the DFG Grants We1748/7 (M.W.), HA1517/31-2 (S.D. and P.H.) and the German Excellence Initiative ``Nanosystems Initiative
Munich (NIM)'' (A.V.P., S.D. and P.H.).


\begin{thebibliography}{1000}
\bibitem{restr} It may happen that the new Floquet eigenspectrum appears as an exact re-normalized copy of the
eigenspectrum of the undriven system, see Ref. \cite{holt3}. However, this  noteworthy effect is rather a
feature of the employed model than being generic.
\bibitem{mobility} The  term 'mobility' is intended to denote the dynamical reaction of a quantum particle to
an applied external static force. Here it bears not the strict meaning of the ratio between the  applied force and the asymptotic velocity
of the particle, as typically assigned  in solid-state and kinetic theories ~\cite{kittel}.
\bibitem{vague} This rather vague definition  gives rise to an immediate question: How to define a relation, $\varepsilon(F_{min})$,
such that when the applied bias is larger than $F_{min}$ all
the avoided crossings  narrower than $\varepsilon$ are
ignored. The resolution of this  challenging issue,
however, is beyond the scope of this work.
\bibitem{singleH} We have also performed experimental studies by using a single-harmonic setup in Eq. (\ref{eq:spatial}) with $V_2 = 0$.
We detected for the BEC cloud a similar mobility response, see in [34].
\end{thebibliography}
\end{document}